# A training process for improving the quality of software projects developed by a practitioner

Cuauhtémoc López-Martín

*Department of Information Systems
Universidad de Guadalajara
Periférico Norte N° 799, C.P.
45100, Zapopan, Jalisco, México
(+52) 33 37-70-33-00 ext. 25717
cuauhtemoc@cucea.udg.mx*

Ali Bou Nassif

*Department of Electrical and
Computer Engineering
University of Sharjah
Sharjah, UAE
anassif@sharjah.ac.ae*

Alain Abran

*Department of Software Engineering
and Information Technology
École de technologie supérieure
Université du Québec, Canada
alain.abran@etsmtl.ca*

**Abstract.**

*Background:* The quality of a software product depends on the quality of the software process followed in developing the product. Therefore, many higher education institutions (HEI) and software organizations have implemented software process improvement (SPI) training courses to improve the software quality.

*Objective:* Because the duration of a course is a concern for HEI and software organizations, we investigate whether the quality of software projects will be improved by reorganizing the activities of the ten assignments of the original personal software process (PSP) course into a modified PSP having fewer assignments (i.e., seven assignments).

*Method:* The assignments were developed by following a modified PSP with fewer assignments but including the phases, forms, standards, and logs suggested in the original PSP. The measurement of the quality of the software assignments was based on defect density.

*Results:* When the activities in the original PSP were reordered into fewer assignments, as practitioners progress through the PSP training, the defect density improved with statistical significance.

*Conclusions:* Our modified PSP could be applied in academy and industrial environments which are concerned in the sense of reducing the PSP training time.

1. Introduction

In 2013, the $407.3 billion software industry [1] was forecast to have a 9.4% compound annual growth rate through 2018 [2]. The quality of the software product is a major ongoing concern in the industry. It relies on the degree of compliance to specified requirements that are related to the features and functions to be developed [3]. A study based on an analysis of 50,000 software projects developed between 2003 and 2012, reported that only 39% of the projects were delivered with the required quality, 43% delivered less than the quality required, and the remaining 18% were cancelled prior to completion or delivered and never used [4]. Another survey based on 8,380 projects and involving 365 information technology executive managers of large, medium, and small companies (from banking, securities, manufacturing, retail, wholesale, heath care, insurance, services, and local, state, and federal organizations), reported that more than a quarter of the software projects were completed with only 25% to 49% of the originally specified features and functions. On average, only 61% of the originally specified features and functions were available at project completion. More specifically, large companies reported that the end product contained 42% of the features and functions, medium companies 65%, and small companies, 74% [5].

Software engineering research aims to improve practice in any of its areas such that research results are useful [6]. The software process is a set of activities, methods, and practices that software engineers and users use to develop and maintain software products [7]. Software process improvement (SPI) aims to understand the software process that is used within an organization and to drive the implementation of changes to that process to achieve objectives such as higher product quality or reduced costs [8].

A systematic literature review of 148 SPI studies published between 1991 and 2008 reported the following findings [9]:

1) Seven distinct evaluation strategies were identified: the most common type, a pre-post comparison (i.e., the SPI strategy was evaluated by comparing the success indicators before and after the SPI strategy took place) was found in 49% of the studies.

2) In 62% of the 148 studies, the process quality was the most measured attribute.

The quality of a software product depends on the quality of the software process used to build the product [10]; hence, software development organizations strive to improve their software processes [10]. Training is one of the most important and reliable human resource techniques to enhance organizations and individual productivity [11]. Furthermore, training in SPI requires an effort that should be addressed three levels of training: the organization, team, and individual [12]. There have been SPI proposals focused on software organizations such as capability maturity model (CMMI) [13], on teams of developers such as team software process (TSP) [14], and at the level of the individual, personal software process (PSP) [15][16]. PSP, which is applied at individual level, was designed to tackle some of the difficulties organizations and teams had in applying CMM practices [17]. CMM has five maturity levels and contains 18 key process areas (KPAs); PSP involves 12 of those KPAs [15].

While there are many claimed benefits for SPI, this study is related to quality improvement at the individual level (in this study, the term *practitioner* is equivalent to *participant*, *software developer*, *software engineer*, and *graduate student*). Other benefits include reductions in the cost of delivering poor quality and software development, and improvements in productivity, on-time delivery, consistency of budget and schedule delivery, customer satisfaction, and employee morale [18].

Software quality starts with the individual software developer: if any of the software project modules developed by a software developer has numerous defects, the modules will be difficult to test and require time to integrate into the system [16]. Our study focuses on the application of SPI specifically designed to improve the quality of a software developer: the PSP, which presents a disciplined process that includes defect management and allows a software developer to produce high-quality software [16].

PSP first scales down industrial software practices to fit the needs of a module-sized software project development, then it guides the software developers through a progressive sequence of practices that provide a foundation for large-scale software development [16].

To analyze the benefits of PSP when used in industrial software projects, Green et al. [19] surveyed 63 software developers who had been trained in PSP and were using it in software development projects. The 63 participants belonged to 24 different organizations and 40% had master degrees. The quantitative part of the study examined whether a perceived gain in software quality is a significant factor in determining the value of PSP to software developers. A seven-point Likert scale was used to investigate responses to the statement "Use of PSP has decreased the number of errors in the software products I build" and Cronbach alpha test was applied to determine the reliability of the scale. After a quantitative analysis, the perceived quality benefits explained 66% of the variance in the perceived usefulness of PSP at a 99% confidence level; therefore, Green et al. [19] encouraged project managers to adopt formal SPI methods to obtain positive impacts on their quality.

The original PSP course training consists of ten assignments [15]. In our study, the PSP practices involved were reordered in accordance with Lopez-Martin and Abran [20] and applied on a smaller number of assignments. Our study was based upon the following observations or conclusions from the literature:

1) Course duration is a principal concern for organizations [16].
2) SPI has often been taught in a few weekly sessions to software development professionals who work in software development and do not have formal education in software engineering [21].
3) The redesign of SPI courses has been encouraged for several years [22].
4) A research topic has been suggested about how the PSP benefits can be obtained with a much smaller training program than the standard PSP course [23].
5) The classroom is a good place to begin acquainting students with the principles of process management in software engineering, and inculcating in students the habit of adhering to these principles as a matter of routine practice [24].
6) An individual disciplined process affects software quality at a 95% confidence level [30], and the later the defects are found, the more costly it is to remove them [25].
7) The PSP, and not repetition of programming assignments, is the most plausible cause of important software quality improvements [26].

The study reported here uses the same process as in a previous study: it is, therefore, necessary to distinguish between *replication* and *reproduction* in a research context. A *replication* of an experiment has been defined as follows [27]:

1) "A conscious and systematic repeat of an original study."
2) "The repetition of an experiment to double-check its results."
3) "A repetition of a research procedure to check the accuracy or truth of the findings reported."

These three definitions imply an explicit relationship with "a previous study" [27]. On one hand, the goal of *replication* in empirical sciences is to test the same hypothesis in a different study [27]. On the other hand, *reproduction* re-examines the results from a previous experiment, using a different experimental protocol [28]. In other words, a *replication* assesses the confidence level for the results of the original experiment to improve the internal validity and reliability of the conclusions, while generalizability is studied by reproduction, improving the external validity [28]. Our study is a *reproduction* since the same process proposed in Lopez-Martin & Abran [20], whose goal was to improve effort prediction, was applied in our study with the goal of improving quality.

The main objective of our study was to demonstrate that as a software developer progresses through the PSP training assignments, the quality of his/her assignments improves, but this can be done with fewer assignments than the ten in the original PSP set. In our study, the phases, reviews, forms, standards, and logs used in the original PSP depicted in Appendix A, have been reordered into a modified software process of seven assignments described in Table 4.

The null ($H_0$) and alternative ($H_1$) hypotheses that were tested in the study and are reported here are the following:

$H_0$: When the activities in the original PSP are reordered into a modified software process having fewer assignments, as practitioners progress through the PSP training, the defect density does not improve with statistical significance.

$H_1$: When the activities in the original PSP are reordered into a modified software process having fewer assignments, as practitioners progress through the PSP training, the defect density improves with statistical significance.

Considering the recommendation by Paulk [29] [30] that the programming language should be taken into account when analyzing software quality when PSP is applied, we set out the following secondary research question:

Does the programming language used influence the quality (defect density) of assignments?

Based upon this question, the hypotheses were also tested taking into account the defect density by two programming languages.

The sample dataset size was 181 practitioners who developed 1,267 software assignments written in C++ and Java. Defect density data of these assignments are included in the Appendix B of this study. The quality of the assignments was measured from the defect density, which was calculated by dividing the number of defects removed by the software assignment size measured in *added* and *modified* 1000 lines of code (KLOC) [15].

The contribution of our study is to investigate whether the quality of individually developed software projects improved by reordering and reducing the assignments in the original PSP.

The rest of this paper is organized as follows: Section 2 briefly describes the phases, standards, logs, and reports in the PSP. Section 3 presents related work on PSP. Section 4 presents the experimental design. Section 5 contains the analysis of quality across assignments. Finally, Section 6 presents a discussion, conclusions, the limitations of the study, and suggestions for future work.

**2. Personal Software Process (PSP)**

The PSP was proposed by Humphrey in 1995 [15] to provide engineers with a disciplined personal framework for developing software projects. From a quality perspective, the goal of PSP is not only to reduce defect density but also to find defects at an earlier stage in the development cycle [25].

An advantage of PSP is that its structure is simple and independent of technology. PSP prescribes no specific languages, tools, or design methods [17] but rather consists of a set of phases, standards, logs, and reports that teach software engineers how to plan, measure, and manage their work. A brief description follows [15]:

Phases: (1) Plan: estimate the size, time, and defects for the project; (2) Design: design the program; (3) Design review; (4) Code: implement the design in any programming language; (5) Code review; (6) Compile: compile the program and fix and log all defects found; (7) Test: execute the program and fix and log all defects found; and (8) Post-mortem: record actual time, defect, and size data on the plan.

Reviews: design review and code review are structured, data-driven processes that are guided by checklists derived from the historical defect data as recorded in the defect recording log.

Forms: a plan summary form is used to document planned and actual results, a test form is used to record data on each of the tests, and a process improvement proposal form is used to record process problems and proposed solutions.

Standards:
1) Defect type – each defect is classified according to documentation (comments, messages), syntax (spelling, punctuation, typos, instruction formats), build and package (change management, library, version control), assignment (declaration, duplicate names, scope, limits), interface (procedure calls and references, I/O, user formats), checking (error messages, inadequate checks), data (structure, content), function (logic, pointers, loops, recursion, computation), system (configuration, timing, memory), environment (design, compile, test, or other support system problems).
2) Coding – a guide to code each assignment. This guide establishes a consistent set of coding practices, provides criteria for judging the quality of the code produced in each assignment, and facilitates the lines of code (LOC) counting. This standard includes how the following issues should be written: header format, identifiers, comments, blank spaces, and indenting [15].
3) Counting – a framework for describing software size measurements.

Logs: (1) A time recording log that tracks the number of minutes software developers spend in each PSP phase, and (2) A defect recording log, which records for each defect the date, sequence number, defect type, phase in which the defect was injected, the phase in which it was removed, the fix time, and a description of the problem.

Regarding the size of assignments, the PSP uses LOC. There are two of them: physical and logical [15]. The counting of physical LOC gives the size in terms of the physical length of the code as it appears when printed. PSP considers *New, Changed,* and *Reused* LOC and all of them were considered as physical LOC for this study. N&C is composed of added (new) and modified (changed) code. The added code is the LOC written during the current programming process, while the modified code is the LOC changed in the base program when modifying a previously developed program. The base program is the total LOC of the previous project while the reused code is the LOC of previously developed assignments that are used without any modification [15].

PSP involves ten assignments distributed in four levels labeled PSP0, PSP1, PSP2, and PSP3. Each of the first three levels (PSP0, PSP1, and PSP2) consists of three assignments and the last (PSP3), of only one. These levels incrementally introduce the set of phases, forms, standards, and logs through ten assignments. PSP0 provides an introduction to the PSP and establishes an initial base of historical size, time, and defect data. PSP1 focus on personal project management techniques, introducing size and effort estimating, schedule planning, and schedule tracking methods. PSP2 adds quality management methods by means of personal design and code reviews, a design notation, design templates, design verification techniques, and measures for managing process and product quality. PSP3 addresses the need to efficiently scale the PSP up to larger projects [48].

In addition to the original PSP of ten assignments, there are two versions of PSP courses consisting of eight (PSPI/II) and seven (PSP Fund/Adv) assignments. These two versions consist of the following six PSP levels: PSP0 describes the current software process, basic collection of time and defect data. PSP0.1 defines a coding standard, basic technique to measure size, and a basic technique to collect process improvement proposals. PSP1 includes techniques to estimate size and effort, and documentation of test results. PSP1.1 involves task and schedule planning. PSP2 includes techniques to review code and design, and PSP2.1 introduces design templates [43].

## 3. Related work

In addition to studies focusing on PSP and quality improvement, other studies show that PSP has been used to develop specific technologies [31] [32] [33] [34]. Moreover, in a number of studies several tools have been proposed for managing PSP [35] [36] [37] [38] [39] [40].

Within the studies specifically related to the quality of software developed following PSP, we searched for those having the following four features:
1) Several software assignments developed by any participant.
2) Involving several participants.
3) Results based upon statistical significance.
4) Application of all practices suggested in the original PSP.

Table 1 shows features of the seven identified studies having these four features, whereas Table 2 presents statistical comparison across PSP quality of studies of Table 1. The conclusions of Rombach et al. [44], Shen et al. [45], Runeson [46], Hayes [47] [48], and Wesslén [25] were based on an overall defect density, whereas that of Paulk [29] [30], and Grazioli & William [43] were based on test defect density.

Grazioli & William [43] used the PSPI/II and PSP Fund/Adv versions. They grouped PSP0 and PSP0.1 in PSP0, PSP1 and PSP1.1 in PSP1, and they analyzed PSP2 and PSP2.1 separately. The assignments by PSP level were distributed as follows [43]:
  a) PSP Fund/Adv

PSP0: first assignment, PSP1: second assignment, PSP2: third and fourth assignments, and PSP2.1: fifth to seventh assignments.
  b) PSPI/II

PSP0 and PSP0.1: first and second assignments, PSP1 and PSP1.1: third and fourth assignments, PSP2: fifth assignment, PSP 2.1: sixth to eight assignments.

The results showed in Table 2 related to Grazioli & William [43] correspond to those obtained in both courses (i.e., the PSPI/II and PSP Fund/Adv versions).

Table 3 shows the programming languages used by each study. In studies on defect density analysis by programming language, Paulk [30] did not find a statistically significant difference when comparing the defect density among the programming languages used (i.e., C, C++, Java, and Visual Basic). Rombach et al. [44] clustered developers by paradigm and did not specify the names of programming languages. Their conclusions illustrate similar trends for each cluster. Shen et al. [45], Runeson [46], Hayes [47], and Wesslén [25] did not report defect density analysis by programming language.

Table 1. Studies involving PSP quality analysis (U: Undergraduate, G: Graduate, "NR" means none reported)

| | Study | | | | | | |
|---|---|---|---|---|---|---|---|
| | Test defect density | | Overall defect density | | | | |
| Topic | Grazioli & William [43] | Paulk [30] | Rombach et al. [44] | Shen et al. [45] | Runeson [46] | Hayes [47] | Wesslen [25] |
| Publication year | 2012 | 2006 | 2008 | 2011 | 2001 | 1998 | 2000 |
| Number of participants | 93 | 1345 | 1636 | 16 | 131 | 181 | 131 |
| Academic level of participants | NR | U, G | U, G | G | G | U, G | G |
| Number of assignments developed by participants | 7,8 | 10 | 10 | 10 | 9 | 9 | 9 |
| Comparison among programming languages used | No | Yes | Yes | No | No | No | No |

Table 2. Statistical comparison across PSP quality studies ($D^{95}$ and $D^{99}$ means a significant difference with $\alpha < 0.05$ and $\alpha < 0.01$, respectively. "NR" means none reported). *PSP2.1 is only used in [43]*

| | Study | | | | | | |
|---|---|---|---|---|---|---|---|
| PSP level comparison | Grazioli & William [43] | Paulk [30] | Rombach et al. [44] | Shen et al. [45] | Runeson [46] | Hayes [47] | Wesslen [25] |
| PSP0 vs. PSP1 | $D^{95}$ | $D^{95}$ | $D^{99}$ | $D^{99}$ | $D^{95}$ | $D^{99}$ | $D^{99}$ |
| PSP1 vs. PSP2 | $D^{95}$ | $D^{95}$ | - | $D^{99}$ | - | - | - |
| PSP2 vs. (PSP3 or PSP2.1) | - | $D^{95}$ | $D^{99}$ | NR | NR | NR | NR |

Table 3. Programming languages used in study ("NR" means none reported)

| Study | Programming languages |
|---|---|
| Grazioli & William [43] | NR |
| Paulk [30] | C, C++, Java, and Visual Basic |
| Rombach et al. [44] | Object-oriented, structured, and other languages |
| Shen et al. [45] | C++, and Java |
| Runeson [46] | C, and Java |
| Hayes [47] | NR |
| Wesslen [25] | Ada, C, C++, Java, Lisp, Pascal, and Simula |

As for PSPI/II and PSP Fund/Adv versions, we only identified one study analyzing data obtained from these two versions of PSP courses [43]; however, we cannot access to raw data analyzed in their study such that we could calculate *overall defects/KLOC*.

We found three additional studies whose conclusions were not based on a statistically significant difference [23] [41] [42] and one study, in which a statistically significant difference was obtained [17]; however, statistical significance analyses involving the defect density values from the second to ninth assignments were not reported [17].

Humphrey [17] compared the quality of software assignments developed by 104 participants, not all of whom worked in software organizations: 80 took PSP in university courses and the rest in organizational courses. Ten assignments were developed by engineers using six different programming languages: Ada, C, C++, Fortran, Pascal, and Visual Basic; however, only Ada, C, and C++ were statistically compared for defect density. Humphrey did not find a statistically significant difference in defect density among the three languages. Results showed that defect density improved, with statistical significance, between the first and tenth assignments with 116.4 defects/KLOC in the first assignment and 48.9 defects/KLOC in the tenth.

Phipps [41] compared individual software assignments developed following PSP using C++ and Java programming languages. Phipps involved only one developer, two software assignments and one assignment by programming language used. In addition, the developer was learning PSP when he developed his first assignment (coded in C++).

Prechelt and Unger [23] compared two groups of undergraduate students: a first group of 24 PSP-trained developers and a second non-trained group of 16. Both groups developed a single software assignment. PSP-trained students were not specifically asked to use PSP techniques. The programming language used by developers in the first group was Java, C, C++, Sather-K, and Modula-2, whereas the second group used Java and C++. The experiment was designed to be independent of the programming language used. Prechelt and Unger concluded that the performance improvements for the PSP-trained group were smaller than the results from PSP proponents usually had assumed, possibly due to the low usage of PSP practices by the PSP trained group.

Ramingwong and Ramingwong [42] compared the defect density of software assignments developed by 13 undergraduate students who followed PSP. Each student developed seven assignments using C++, PHP, Java, and Visual Basic. Comparison was based on an average defect density by assignment. The averages reported from the second to seventh assignment were 97, 70, 69, 60, 98, and 63 defects/KLOC, respectively.

### 4. Experimental design

The 181 software developers in our study were registered in a university master degree program (all of them had gained a bachelor degree). The course was elective (or optional) and it was taught in either public or private universities. In the private ones, the participants paid for the course as part of their set of semester subjects. Graduate students were selected because the original PSP training course was aimed at graduate software engineers who had the required programming language proficiency and software development competence [17]. The course was part of a semester subject, the duration by course was dependent of the university (from twelve to sixteen weeks), one day per week was assigned to the course, and four hours were allocated per day (one lecture-hour and three practical-hours). Introduction to PSP, code and counting standards were taught in the first day. One assignment was performed daily from the second to the eighth day. The ninth day was allocated to the final report, and the tenth day to analyze the data obtained from all assignments of developers. The rest of the semester-course was allocated to theoretical topics related to software processes and their statistical analysis.

The 1,267 software assignments in our study were developed between 2005 and the second semester of 2012. Each student selected his own programming language. The 1,267 assignments were coded in C++ and Java. They were selected because they corresponded to the two larger data sets with 686 (Java) and 581 (C++) assignments (see Table 5).

The measurement of the quality of the software assignments was based on defect density calculated by dividing the number of defects removed by the program size measured in KLOC [15]. This normalization offsets the size differences among assignments [15]. Three measures are typically proposed to study the effect of PSP practices on the defect density: the overall, compile, and test [15]. Our study was related to the overall defect density as reported in [25] [44] [45] [46] [47] [48]. Furthermore, to determine the quality of the development process, it is recommended to use the *added* and *modified* LOC for calculating the defect density [15], not taking *reused* lines of code into account: i.e., in our study *defect density = overall defects/KLOC*, where *KLOC* means *added* and *modified* KLOC.

In our study, only data of participants who followed all the phases, forms, standards, and logs suggested in the original PSP [15] were selected. Those ones who did not follow them could continue the course; however, their data were not considered for our study. Eighteen developers were excluded because they did not follow the correct process. Therefore, since participants were not selected randomly, the experimental design was considered quasi-experimental [49].

Our study was based on a modified PSP with fewer assignments but it included the phases, forms, standards, and logs suggested in the original ten assignments. Table 4 describes them in terms of the sequence number of the assignment in which a given activity is used. In Table 4, expert judgment estimation refers to that technique based on intuition and derived of the experience of practitioners on similar projects [20]; moreover, with PROBE (PROxy-based estimating), developers used the relative size of a proxy to make their initial estimate, then used historical data to convert the relative size of the proxy to LOC [15].

Another exception is the generation of a simple linear regression (SLR) in assignment six from a regression analysis involving more pairs of data: five assignments instead of three as in the original PSP. This SLR is used for predicting the development effort and it includes to *N&C* lines of code as independent variable and effort as dependent variable. The SLR analysis is based on identification and exclusion of outliers from scatter plots, as well as the interpretation of correlation ($r$) and determination ($r^2$) coefficients. The identification of outliers was based on observations which had either Studentized residuals greater than 2 in absolute value, or leverage values greater than 3 times that of an average data point. Besides of these two cases, an assignment was only excluded when a non-statistical specific reason was identified by the practitioner, such as by using a function or a library not used before which spent more effort. A SLR was used when its $r^2 \geq 0.5$ [15].

The effort prediction using a multiple linear regression (MLR) in assignment seven represents a new proposal (it is not applied for predicting the effort of any of the assignments in the original PSP). The MLR includes the analysis of *N&C* and *Reused* lines of code as independent variables, and *effort* as dependent variable. The MLR is generated from the data analysis of six assignments.

The experiment was performed in a controlled environment with the following characteristics [20]:
- All participants were working on software development in organizations, and none of them had previously taken a course related to PSP.
- All the participants were registered in a postgraduate program in computer science.
- Participation in the study was not mandatory, and the participants did not receive any payment for attending the course.
- Each participant selected the programming language he/she wanted to use in the assignments, based on his/her expertise. They were warned that the PSP course was not suitable to simultaneously learn a new programming language.
- Participants had already taken at least one course on the programming language used in the assignments.
- To reduce bias, participants were neither informed of our experimental objective, nor penalized for their performance regarding quality of their assignments (i.e., in the sense of obtaining a lower or higher defect density). They were penalized if they did not follow the guidelines of the PSP process.
- With the goal of practicing the data records on forms and logs, participants manually filled in a spreadsheet for each assignment instead of using any PSP tool. This spreadsheet was submitted for review.
- There were no more than 15 participants in each course.

- Participants were supervised and mentored on the process during the assignments: after each assignment, their documentation was reviewed and they received feedback when requested, about any issue before starting their next assignment.
- Only one assignment was performed by day as suggested in the original PSP course.
- All the participants adopted a coding and counting standard.
- The code written in each assignment was designed by the participant such that it could be reused in subsequent assignments.
- The assignments had a complexity similar to that suggested in [15]. In each course, from a set of 18 assignments, a subset of seven was randomly assigned to each of the participants.

Data from 1,575 software assignments developed by 225 software developers were gathered from 2005 to the second semester of 2012. Table 5 presents the number of participants and software assignments by programming language used. Of the assignments in Table 5, only those coded in Java or in C++ were selected for study because of the large number of software assignments.

Regarding the secondary research question written in the Introduction section, the following hypotheses to be tested were formulated:

$H_{0CJ}$: There is not a statistically significant difference in the defect density between the assignments coded in C++ and those coded in Java at a 95% confidence level.

$H_{1CJ}$: There is a statistically significant difference in the defect density between the assignments coded in C++ and those coded in Java at a 95% confidence level.

These two hypotheses are formulated because in our study, the developers used an Integrated Development Environments (IDEs) which had features such as automatic code completion, and automatic compilation of individual Java statements. That is, in our study, Java developers did not record defects in the compile phase.

If $H_{0CJ}$ is rejected, defect density analysis should be carried out separately for C++ and Java assignments. Otherwise, the assignments can be pooled for the quality analysis.

Table 6 presents the adopted counting standard for C++ and Java lines of code.

Table 4. Description of the modified PSP for this study
(a "√" means a given activity used in the assignment)

| | | Assignment | | | | | | |
|---|---|---|---|---|---|---|---|---|
| | | First | Second | Third | Fourth | Fifth | Sixth | Seventh |
| Phases | Plan | | | | | | | |
| | Total N&C size planning | √ | √ | √ | √ | √ | √ | √ |
| | Time estimation from expert judgment | √ | √ | √ | √ | √ | √ | √ |
| | N&C, reused, and deleted LOC size planning | | √ | √ | √ | √ | √ | √ |
| | Defect planning | | | | | √ | √ | √ |
| | PROBE method | | | | | | √ | √ |
| | Effort estimation from simple linear regression | | | | | | √ | √ |
| | Effort estimation from multiple linear regression | | | | | | | √ |
| | Design | √ | √ | √ | √ | √ | √ | √ |
| | Design review | | | | √ | √ | √ | √ |
| | Code | √ | √ | √ | √ | √ | √ | √ |
| | Code review | | | | √ | √ | √ | √ |
| | Compile | √ | √ | √ | √ | √ | √ | √ |
| | Testing | √ | √ | √ | √ | √ | √ | √ |
| | Postmortem | √ | √ | √ | √ | √ | √ | √ |
| Reviews | Code review checklist | | | | √ | √ | √ | √ | √ |
| | Design review checklist | | | | | √ | √ | √ | √ |
| Forms | Project plan summary | √ | √ | √ | √ | √ | √ | √ |
| | Process improvement proposal | √ | √ | √ | √ | √ | √ | √ |
| | Test report template | | | √ | √ | √ | √ | √ |
| Standards | Defect type | √ | √ | √ | √ | √ | √ | √ |
| | Coding standard | √ | √ | √ | √ | √ | √ | √ |
| | LOC counting standard | √ | √ | √ | √ | √ | √ | √ |
| Logs | Time recording log | √ | √ | √ | √ | √ | √ | √ |
| | Defect recording log | √ | √ | √ | √ | √ | √ | √ |

Table 5. Number of assignments by programming language

| Language | Participants | Assignments |
|---|---|---|
| Java | 98 | 686 |
| C++ | 83 | 581 |
| Visual Basic | 28 | 196 |
| Delphi | 7 | 49 |
| PHP | 6 | 42 |
| ABAP | 1 | 7 |
| Visual FoxPro | 1 | 7 |
| Perl | 1 | 7 |

Table 6. Counting standard for C++ and Java lines of code

| | |
|---|---|
| 1) Count type | |
|     Physical/logical | Physical |
| 2) Statement type | Included |
| 3) Executable | Yes |
| 4) No executable | |
|     Declarations | Yes, one by text line |
|     Compiler directives | Yes, one by text line |
|     Comments | No |
|     Blank lines | No |
| 5) Clarifications | |
|     *{* and *}* | Yes |

## 5. Quality analysis of assignments

The $H_{0CJ}$ and $H_{1CJ}$ hypotheses on defect density were tested with 686 and 581 projects coded in Java and C++, respectively, by comparing the two sets of assignments (one set by programming language), assignment by assignment. Table 7 presents the mean and median values of defect density for each assignment.

A suitable statistical test was selected, taking into account the number and size of sets to be compared, as well as the dependence, normality and variance of the data [50]. The two sets (each corresponding to a programming language) were independent of each other as each set of developers, made up of individual software developers, developed their own assignments. Table 8 shows the normality statistical analysis by programming language for the seven assignments.

Table 7. Defect density by programming language

| Assignment | Mean | | Median | |
|---|---|---|---|---|
| | C++ | Java | C++ | Java |
| First | 148 | 89 | 125 | 74 |
| Second | 125 | 88 | 115 | 68 |
| Third | 99 | 75 | 85 | 49 |
| Fourth | 95 | 52 | 70 | 36 |
| Fifth | 79 | 43 | 65 | 31 |
| Sixth | 59 | 32 | 48 | 18 |
| Seventh | 53 | 32 | 40 | 0 |

Table 8. P-values of normality tests by assignment for C++ and Java

| Normality test | Assignment | | | | | | | | | | | | | |
|---|---|---|---|---|---|---|---|---|---|---|---|---|---|---|
| | First | | Second | | Third | | Fourth | | Fifth | | Sixth | | Seventh | |
| | C++ | Java | C++ | Java | C++ | Java | C++ | Java | C++ | Java | C++ | Java | C++ | Java |
| Chi-squared | 0.0928 | 0.0160 | 0.0822 | 0.0000 | 0.0000 | 0.0000 | 0.0001 | 0.0000 | 0.0035 | 0.0000 | 0.0000 | 0.0000 | 0.0000 | 0.0000 |
| Shapiro-Wilk | 0.0003 | 0.0000 | 0.0006 | 0.0000 | 0.0000 | 0.0000 | 0.0000 | 0.0000 | 0.0000 | 0.0000 | 0.0000 | 0.0000 | 0.0000 | 0.0000 |
| Skewness | 0.0245 | 0.0250 | 0.1322 | 0.0192 | 0.0141 | 0.0003 | 0.0015 | 0.0000 | 0.0045 | 0.0197 | 0.0000 | 0.0003 | 0.0004 | 0.0000 |
| Kurtosis | 0.2231 | 0.9629 | 0.2480 | 0.5545 | 0.2763 | 0.0043 | 0.0138 | 0.0000 | 0.0792 | 0.8193 | 0.0000 | 0.0016 | 0.0001 | 0.0000 |

As for variance data, Table 9 shows results of the Levene test which compared the two sets of software projects by programming language. The Levene test tests the null hypothesis that the standard deviations of defect density within each of the two programming languages are the same. Figure 1 shows the residual plot for the first assignment. From the p-values in Table 9, it was concluded that there are not statistically significant differences between the standard deviations at the 99% confidence level from the second to seventh assignments, and that there is statistically significant difference between the standard deviations at the 99% for the first assignment.

Table 9. Levene test between C++ and Java sets grouped by assignment

| Assignment | p-value |
|---|---|
| First | 0.0011 |
| Second | 0.2067 |
| Third | 0.7191 |
| Fourth | 0.0291 |
| Fifth | 0.0210 |
| Sixth | 0.0492 |
| Seventh | 0.0556 |

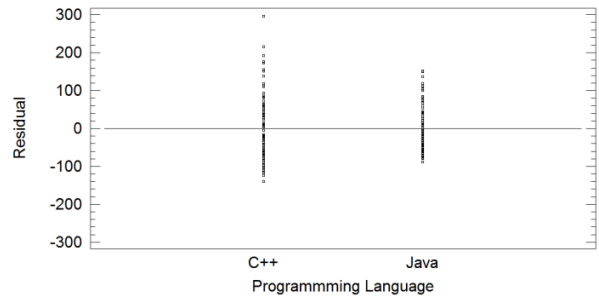

Figure 1. First assignment residual plot for defect density

As two sets of projects were compared (corresponding to the two programming languages), the two sets were independent, the size of them are different (i.e., 98 vs. 83 assignments), the first assignment skewness p-values for C++ and Java sets are greater than 0.01 (Table 8), that is, both sets are symmetric with 99% confidence, there are not differences between the variances from the second to seventh assignments (Table 9), and since a *t-test* statistical test is sufficiently robust, except when skew is severe or when variances and data set sizes both differ [50], a *t-test*, which tests the null hypothesis that the mean of defect density within each of the two programming languages are the same, was used to compare the two sets by assignments. Table 10 shows the p-values for each assignment. It shows that there was a statistically significant difference between the mean at 99% confidence level for five sets of assignments, 95% confidence for the seventh assignment, and 90% confidence for the third assignment. As a graphical example, Figure 2 shows a box-and-whisker for C++ and Java for the first assignment.

Based on the results presented in Table 7 and Table 10, the following hypothesis for six of the seven assignments was accepted (the third assignment was accepted with 90% confidence level):

$H_{1CJ}$: There is a statistically significant difference in the defect density between the assignments coded in C++ and those coded in Java at a 95% confidence level.

Table 10. Defect density statistical comparison between C++ and Java by assignment

| Assignment | *t-test* p-value |
|---|---|
| First | 0.0000 |
| Second | 0.0009 |
| Third | 0.0747 |
| Fourth | 0.0001 |
| Fifth | 0.0000 |
| Sixth | 0.0005 |
| Seventh | 0.0133 |

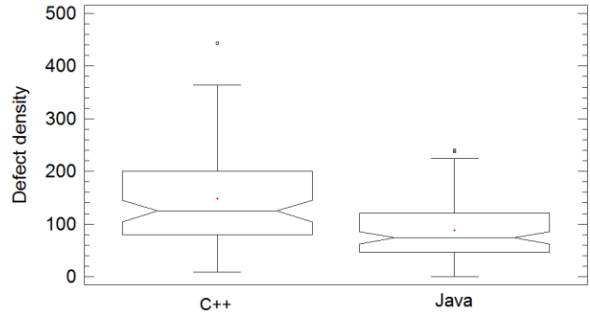
Figure 2. First assignment (from Table 10) box-and-whisker plot for C++ and Java assignments

Since $H_{0CJ}$ was rejected, the defect density analysis should be carried out by individual programming language for C++ and Java software projects. Therefore, the following additional hypotheses were derived from those formulated in the introduction section of our study:

$H_{0C++}$: When the activities in the original PSP are reordered into a modified software process having fewer C++ assignments, as practitioners progress through the PSP training, the defect density does not improve with statistical significance.

$H_{1C++}$: When the activities in the original PSP are reordered into a modified software process having fewer C++ assignments, as practitioners progress through the PSP training, the defect density improves with statistical significance.

$H_{0Java}$: When the activities in the original PSP are reordered into a modified software process having fewer Java assignments, as practitioners progress through the PSP training, the defect density does not improve with statistical significance.

$H_{1Java}$: When the activities in the original PSP are reordered into a modified software process having fewer Java assignments, as practitioners progress through the PSP training, the defect density improves with statistical significance.

There are seven assignments that each software developer made: that is, each pair of data sets to be compared is dependent (also termed *related* or *paired*). Therefore, in addition to the number of sets to be compared and dependence of data, a suitable statistical test to compare the defect density of assignments was selected taking into account a normality analysis of the set of defect density differences by pair of assignments: if these set of differences were normally distributed, then a *t-paired* statistical test was used, otherwise, a Wilcoxon test was applied.

Table 11 shows the normality test that had the smallest p-value among the kurtosis, skewness, chi-squared, and Shapiro-Wilk normality tests by pair of assignments: if this p-value is greater than or equal to 0.01, we cannot reject the idea that the set of differences comes from a normal distribution with 99% confidence, otherwise we can reject the idea that the set of differences comes from a normal distribution with 99% confidence. From Table 11 we can interpret that the *t-paired* test had to be applied in six of the 21 pairs for C++, and only two cases for Java.

Results from Table 7 and Table 12 allowed us to accept the following hypotheses (except for three pairs related to C++: Third – Fourth, Fourth – Fifth, and Sixth – Seventh, and for three Java cases: First – Second, Fourth – Fifth, and Sixth – Seventh):

$H_{1C++}$: When the activities in the original PSP are reordered into a modified software process having fewer C++ assignments, as practitioners progress through the PSP training, the defect density improves with statistical significance.

$H_{1Java}$: When the activities in the original PSP are reordered into a modified software process having fewer Java assignments, as practitioners progress through the PSP training, the defect density improves with statistical significance.

Table 11. Smallest p-value among normality tests by pair of assignments for C++ and Java

| Pair | C++ | | Java | |
|---|---|---|---|---|
| | Normality test | p-value | Normality test | p-value |
| First – Second | Kurtosis | 0.1118 | Chi-Squared | 0.0767 |
| First – Third | Kurtosis | 0.1451 | Shapiro-Wilk | 0.0003 |
| First – Fourth | Chi-Squared | 0.2730 | Shapiro-Wilk | 0.0000 |
| First – Fifth | Kurtosis | 0.2795 | Shapiro-Wilk | 0.0047 |
| First – Sixth | Skewness | 0.1283 | Chi-Squared | 0.0856 |
| First – Seventh | Kurtosis | 0.0082 | Kurtosis | 0.0000 |
| Second – Third | Kurtosis | 0.0088 | Chi-Squared | 0.0001 |
| Second – Fourth | Chi-Squared | 0.0002 | Shapiro-Wilk | 0.0000 |
| Second – Fifth | Chi-Squared | 0.1176 | Shapiro-Wilk | 0.0000 |
| Second – Sixth | Chi-Squared | 0.3880 | Shapiro-Wilk | 0.0018 |
| Second – Seventh | Kurtosis | 0.0873 | Chi-Squared | 0.0003 |
| Third – Fourth | Kurtosis | 0.0593 | Chi-Squared | 0.0000 |
| Third – Fifth | Skewness | 0.3093 | Chi-Squared | 0.0000 |
| Third – Sixth | Kurtosis | 0.0604 | Chi-Squared | 0.0000 |
| Third – Seventh | Shapiro-Wilk | 0.2363 | Chi-Squared | 0.0000 |
| Fourth – Fifth | Chi-Squared | 0.0000 | Shapiro-Wilk | 0.0000 |
| Fourth – Sixth | Shapiro-Wilk | 0.0213 | Shapiro-Wilk | 0.0000 |
| Fourth – Seventh | Shapiro-Wilk | 0.0018 | Chi-Squared | 0.0000 |
| Fifth – Sixth | Chi-Squared | 0.0251 | Chi-Squared | 0.0000 |
| Fifth – Seventh | Kurtosis | 0.0264 | Chi-Squared | 0.0000 |
| Sixth – Seventh | Chi-Squared | 0.0002 | Chi-Squared | 0.0000 |

Results reported in previous PSP studies (see Table 2) were based on four groups of assignments named PSP levels (each of the first three levels consisted of three assignments, while the last level had only one); hence, the assignments of our study were also grouped into four sets of assignments with the goal of comparing our results with the studies in Table 2. The first three groups of our study include two assignments, and the fourth group the seventh assignment. The second group includes the design and code reviews introduced in the third and fourth assignments of our study (the PSP2 level of Table 2 also includes these two reviews), which allows analysing the influence of these two reviews on the quality of assignments. Based on this, the following additional hypotheses were derived:

$H_{0GC++}$: When the activities in the original PSP are reordered into a modified software process having fewer C++ grouped assignments, as practitioners progress through the PSP training, the defect density does not improve with statistical significance.

$H_{1GC++}$: When the activities in the original PSP are reordered into a modified software process having fewer C++ grouped assignments, as practitioners progress through the PSP training, the defect density improves with statistical significance.

$H_{0GJava}$: When the activities in the original PSP are reordered into a modified software process having fewer Java grouped assignments, as practitioners progress through the PSP training, the defect density does not improve with statistical significance.

$H_{1GJava}$: When the activities in the original PSP are reordered into a modified software process having fewer Java grouped assignments, as practitioners progress through the PSP training, the defect density improves with statistical significance.

Table 13 includes the mean and median values for the four groups by programming language. A *repeated measures ANOVA* test for defect density among PSP groups (which is a test for more than two dependent data sets) had a p-value equal to 0.0000 for C++ and Java, which signifies that there was a statistically significant difference among the four groups for the two programming languages.

Table 14 shows the results of a Wilcoxon test (which is a test for two dependent data sets) that involved pairs of PSP groups. In each case, the p-values were less than 0.01; therefore, there was a statistically significant difference at the 99% confidence level. Figures 3 and 4 show box-and-whisker plots, including a median notch, for the C++ and Java programming languages, respectively.

Table 12. Wilcoxon and *t-paired* p-values by pairs of assignments for C++ and Java ($D^{90}$, $D^{95}$ and $D^{99}$ means a significant difference with α ≤ 0.10, α ≤ 0.05 and α ≤ 0.01, respectively)

| Pair | C++ | | Java | |
|---|---|---|---|---|
| | p-value | Difference | p-value | Difference |
| First – Second | 0.0308 | $D^{95}$ | 0.9817 | – |
| First – Third | 0.0002 | $D^{99}$ | 0.0088 | $D^{99}$ |
| First – Fourth | 0.0000 | $D^{99}$ | 0.0000 | $D^{99}$ |
| First – Fifth | 0.0000 | $D^{99}$ | 0.0000 | $D^{99}$ |
| First – Sixth | 0.0000 | $D^{99}$ | 0.0000 | $D^{99}$ |
| First – Seventh | 0.0000 | $D^{99}$ | 0.0000 | $D^{99}$ |
| Second – Third | 0.0168 | $D^{95}$ | 0.0513 | $D^{90}$ |
| Second – Fourth | 0.0042 | $D^{99}$ | 0.0000 | $D^{99}$ |
| Second – Fifth | 0.0000 | $D^{99}$ | 0.0000 | $D^{99}$ |
| Second – Sixth | 0.0000 | $D^{99}$ | 0.0000 | $D^{99}$ |
| Second – Seventh | 0.0000 | $D^{99}$ | 0.0000 | $D^{99}$ |
| Third – Fourth | 0.6293 | - | 0.0059 | $D^{99}$ |
| Third – Fifth | 0.0601 | $D^{90}$ | 0.0016 | $D^{99}$ |
| Third – Sixth | 0.0000 | $D^{99}$ | 0.0000 | $D^{99}$ |
| Third – Seventh | 0.0000 | $D^{99}$ | 0.0000 | $D^{99}$ |
| Fourth – Fifth | 0.1811 | – | 0.3824 | – |
| Fourth – Sixth | 0.0002 | $D^{99}$ | 0.0010 | $D^{99}$ |
| Fourth – Seventh | 0.0000 | $D^{99}$ | 0.0003 | $D^{99}$ |
| Fifth – Sixth | 0.0205 | $D^{95}$ | 0.0011 | $D^{99}$ |
| Fifth – Seventh | 0.0009 | $D^{99}$ | 0.0010 | $D^{99}$ |
| Sixth – Seventh | 0.2355 | – | 0.5377 | – |

Table 13. Defect density of PSP groups by programming language

| PSP groups | Mean | | Median | |
|---|---|---|---|---|
| | C++ | Java | C++ | Java |
| First – Second | 137 | 88 | 121 | 71 |
| Third – Fourth | 97 | 64 | 75 | 44 |
| Fifth – Sixth | 69 | 37 | 52 | 26 |
| Seventh | 53 | 32 | 40 | 0 |

Table 14. Statistical comparison ($D^{99}$ means a significant difference with α ≤ 0.01).

| PSP groups | Programming language | |
|---|---|---|
| | C++ | Java |
| First – Second vs. Third – Fourth | $D^{99}$ | $D^{99}$ |
| Third – Fourth vs. Fifth – Sixth | $D^{99}$ | $D^{99}$ |
| Fifth – Sixth vs. Seventh | $D^{99}$ | $D^{99}$ |

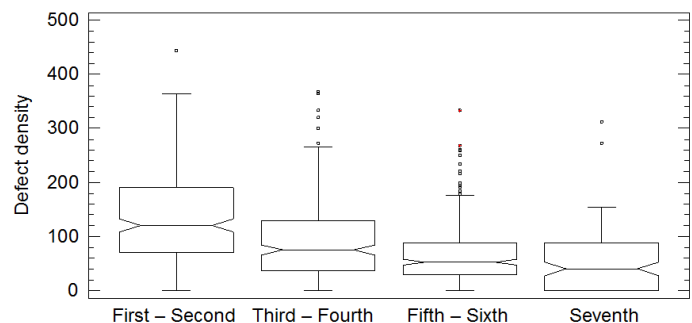

Figure 3. Box-and-whisker plot for C++ software assignments

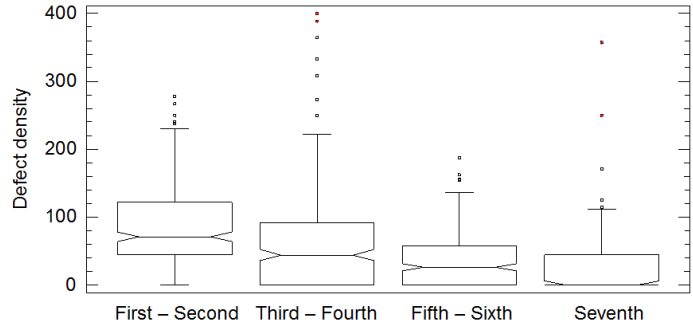

Figure 4. Box-and-whisker plot for Java software assignments

## 6. Discussion

The quality of software is a significant concern for organizations. Recent reports reveal that a low percentage of projects are actually delivered with the required quality.

An aim of software engineering is to deliver software of high quality, and the SPI is a recommended approach to improve software processes and produce high-quality software [8], which can be implemented at the individual, team, or organizational level. The goal of PSP is to provide software engineers the needed training/expertise required to deliver high quality products [15] [16]. Training is one of the most important and reliable human resource techniques to enhance organization and individual productivity [11] and the quality of a software product begins at the individual process level. Training duration has been, however, an industry and HEI concern and in this study we proposed a process containing the same activities as the original PSP but with fewer numbers of assignments (i.e., reduced from ten to seven).

The dataset included 181 practitioners who each developed seven PSP assignments following the same software process, for a total of 1,267 assignments, which were separated for analysis into those coded in C++ and those in Java. The 1,267 assignments, separated by programming language, presented a statistical difference in defect density.

In accordance with Tables 7 and 12, the following hypotheses were accepted (except for three pairs related to C++, and for three Java pairs):

$H_{1C++}$: When the activities in the original PSP are reordered into a modified software process having fewer C++ assignments, as practitioners progress through the PSP training, the defect density improves with statistical significance.

$H_{1Java}$: When the activities in the original PSP are reordered into a modified software process having fewer Java assignments, as practitioners progress through the PSP training, the defect density improves with statistical significance.

Once the assignments were grouped, according to Tables 13 and 14, the hypotheses accepted in our research were the following:

$H_{1GC++}$: When the activities in the original PSP are reordered into a modified software process having fewer C++ grouped assignments, as practitioners progress through the PSP training, the defect density improves with statistical significance.

$H_{1GJava}$: When the activities in the original PSP are reordered into a modified software process having fewer Java grouped assignments, as practitioners progress through the PSP training, the defect density improves with statistical significance.

Table 14 also allowed observing the favorable influence with statistical significance, of the design and code reviews on quality by grouping the third and fourth assignments.

The aim of a course is to improve the skills of students. This is, however, never guaranteed, and it must be verified whether or not the aim has been achieved. There are three possible outcomes: skills not modified, skills improved, and skills degraded. This study has confirmed that indeed the skills have improved by using the defect density as criterion.

The main objective of our study was achieved: we demonstrated that as a software developer progressed through PSP training assignments, software quality improved.

This result is also relevant for the software engineering academic and industrial community because (1) it suggests that software engineering development practices such as plan, design, design review, code, code review, testing, and postmortem are related to software quality improvement, and (2) academy and industrial environments want to reduce the PSP training time.

Six of the seven studies in Table 1 applied the original PSP for analyzing the quality performance across the assignments: three involved graduate and undergraduate students [30] [44] [47], and three only graduate students [25] [45] [46]. In comparison, our study involved only students registered in a graduate course. Only two of the previous six studies analyzed performance by taking into account the programming language: Paulk [30] did not find a statistically significant difference in defect density among the programming languages used (C, C++, Java, and Visual Basic), while Rombach et al. [44], in spite of having clustered the software projects by programming language paradigm (object-oriented, structured, and other languages), concluded that defect density had similar trends for each cluster. In comparing Table 2 versus Table 14, only one of the seven studies [30] achieved a statistical result similar to ours. This study was based on test defect density [30], whereas ours was based on overall defect density.

The limitations of our study include the following: the defects analyzed were pooled, that is, they were not analyzed by defect types nor by the phase in which they were injected (plan, design, design review, code, code review, compile, testing). In addition, our study involved only two programming languages.

A first validity threat of our experiment is related to the use of a spreadsheet instead of a PSP software tool for registering data of logs and forms. It could be a threat since a previous study concluded the data quality problems could be presented when data are manually collected [51]. Based upon this precedent, we controlled that the participants correctly filled their forms and logs. Although all practitioners adopted a same counting standard for C++ and Java lines of code, a second validity threat is that each practitioner designed its own coding standard. A third validity threat could be related to the randomly assigned assignments to each of the developers, making our results less comparable.

In future work, we plan to investigate whether the productivity of developers (LOC/hour) improves by applying the process used in this study; moreover, it would be interesting to develop software tools that help improve the quality of individually developed software assignments, as well as identify the types of defects frequently injected as participants develop their software assignments.

## Acknowledgments

The authors thank the CUCEA of Universidad de Guadalajara, México, and the Consejo Nacional de Ciencia y Tecnología (CONACyT), the University of Sharjah, and the Ecole de technologie supérieure - Université du Québec (Canada), for their support during the development of this work. This work has been partially funded by the National Research Council of Canada.

**Appendix A:** Description of the original PSP (a "√" means a given activity used in the assignment; LOC: lines of code; N&C: new and changed LOC)

|  |  | Assignment |  |  |  |  |  |  |  |  |  |
|---|---|---|---|---|---|---|---|---|---|---|---|
|  |  | First | Second | Third | Fourth | Fifth | Sixth | Seventh | Eight | Ninth | Tenth |
| Phases | Plan |  |  |  |  |  |  |  |  |  |  |
|  | Total N&C size planning |  | √ | √ | √ | √ | √ | √ | √ | √ | √ |
|  | Time estimation from expert judgment | √ | √ | √ | √ | √ | √ | √ | √ | √ | √ |
|  | N&C, reused, and deleted LOC size planning |  |  |  | √ | √ | √ | √ | √ | √ | √ |
|  | Defect planning |  |  |  |  |  |  | √ | √ | √ | √ |
|  | PROBE method |  |  |  | √ | √ | √ | √ | √ | √ | √ |
|  | Effort estimation from simple linear regression |  |  |  | √ | √ | √ | √ | √ | √ | √ |
|  | Design | √ | √ | √ | √ | √ | √ | √ | √ | √ | √ |
|  | Design review |  |  |  |  |  |  | √ | √ | √ | √ |
|  | Code | √ | √ | √ | √ | √ | √ | √ | √ | √ | √ |
|  | Code review |  |  |  |  |  |  | √ | √ | √ | √ |
|  | Compile | √ | √ | √ | √ | √ | √ | √ | √ | √ | √ |
|  | Testing | √ | √ | √ | √ | √ | √ | √ | √ | √ | √ |
|  | Postmortem |  | √ | √ | √ | √ | √ | √ | √ | √ | √ |
| Reviews | Code review checklist |  |  |  |  |  |  | √ | √ | √ | √ |
|  | Design review checklist |  |  |  |  |  |  | √ | √ | √ | √ |
| Forms | Project plan summary | √ | √ | √ | √ | √ | √ | √ | √ | √ | √ |
|  | Process improvement proposal |  | √ | √ | √ | √ | √ | √ | √ | √ | √ |
|  | Test report template |  |  |  | √ | √ | √ | √ | √ | √ | √ |
| Standards | Defect type | √ | √ | √ | √ | √ | √ | √ | √ | √ | √ |
|  | Coding standard |  | √ | √ | √ | √ | √ | √ | √ | √ | √ |
|  | LOC counting standard |  | √ | √ | √ | √ | √ | √ | √ | √ | √ |
| Logs | Time recording log | √ | √ | √ | √ | √ | √ | √ | √ | √ | √ |
|  | Defect recording log | √ | √ | √ | √ | √ | √ | √ | √ | √ | √ |

**Appendix B:** Defect density by assignment by programming language

| C++ | | | | | | | Java | | | | | | |
|---|---|---|---|---|---|---|---|---|---|---|---|---|---|
| First | Second | Third | Fourh | Fifth | Sixth | Seventh | First | Second | Third | Fourh | Fifth | Sixth | Seventh |
| 258 | 222 | 182 | 129 | 143 | 333 | 143 | 173 | 80 | 88 | 47 | 38 | 20 | 0 |
| 85 | 83 | 53 | 33 | 37 | 0 | 0 | 33 | 91 | 0 | 0 | 43 | 0 | 0 |
| 125 | 190 | 87 | 172 | 60 | 53 | 273 | 26 | 20 | 47 | 32 | 17 | 29 | 23 |
| 200 | 94 | 69 | 83 | 67 | 69 | 0 | 71 | 63 | 0 | 0 | 0 | 0 | 0 |
| 231 | 133 | 0 | 56 | 222 | 63 | 143 | 129 | 147 | 128 | 167 | 162 | 154 | 171 |
| 128 | 167 | 91 | 154 | 54 | 41 | 95 | 75 | 152 | 389 | 400 | 52 | 26 | 51 |
| 122 | 267 | 0 | 0 | 56 | 34 | 53 | 68 | 56 | 0 | 59 | 24 | 34 | 38 |
| 159 | 217 | 250 | 42 | 37 | 23 | 0 | 162 | 59 | 0 | 0 | 40 | 0 | 0 |
| 63 | 64 | 125 | 29 | 40 | 57 | 100 | 200 | 188 | 364 | 222 | 116 | 71 | 250 |
| 32 | 111 | 100 | 0 | 0 | 0 | 0 | 122 | 0 | 0 | 0 | 0 | 0 | 0 |
| 120 | 52 | 26 | 65 | 11 | 7 | 11 | 93 | 45 | 0 | 0 | 0 | 0 | 0 |
| 364 | 231 | 250 | 367 | 192 | 200 | 143 | 53 | 0 | 67 | 0 | 25 | 59 | 26 |
| 182 | 38 | 59 | 56 | 59 | 17 | 48 | 32 | 46 | 32 | 0 | 0 | 6 | 29 |
| 59 | 49 | 25 | 23 | 24 | 113 | 68 | 100 | 105 | 104 | 43 | 74 | 83 | 0 |
| 143 | 111 | 85 | 104 | 89 | 62 | 25 | 52 | 68 | 38 | 23 | 11 | 0 | 0 |
| 444 | 114 | 77 | 167 | 179 | 63 | 52 | 77 | 53 | 100 | 59 | 95 | 111 | 45 |
| 303 | 211 | 111 | 118 | 38 | 56 | 83 | 38 | 0 | 0 | 37 | 0 | 0 | 0 |
| 172 | 135 | 333 | 125 | 111 | 250 | 65 | 53 | 0 | 45 | 59 | 26 | 15 | 38 |
| 80 | 57 | 107 | 71 | 102 | 88 | 89 | 115 | 63 | 91 | 100 | 100 | 88 | 100 |
| 211 | 146 | 250 | 91 | 78 | 32 | 97 | 127 | 120 | 59 | 11 | 19 | 13 | 20 |
| 132 | 250 | 182 | 56 | 133 | 50 | 27 | 44 | 57 | 111 | 45 | 130 | 47 | 0 |
| 208 | 45 | 56 | 50 | 34 | 45 | 0 | 32 | 143 | 0 | 67 | 0 | 0 | 0 |
| 121 | 147 | 267 | 105 | 93 | 148 | 111 | 240 | 77 | 0 | 34 | 91 | 25 | 125 |
| 152 | 167 | 50 | 67 | 125 | 38 | 0 | 22 | 0 | 0 | 0 | 27 | 0 | 0 |
| 91 | 148 | 125 | 20 | 0 | 50 | 67 | 100 | 104 | 250 | 100 | 67 | 50 | 57 |
| 161 | 83 | 120 | 217 | 74 | 29 | 125 | 77 | 0 | 50 | 42 | 38 | 0 | 0 |
| 111 | 222 | 143 | 129 | 176 | 54 | 0 | 111 | 148 | 82 | 54 | 56 | 50 | 125 |
| 175 | 79 | 91 | 27 | 14 | 51 | 19 | 200 | 250 | 105 | 53 | 54 | 32 | 83 |
| 233 | 65 | 94 | 108 | 125 | 41 | 33 | 109 | 250 | 63 | 63 | 0 | 56 | 71 |
| 125 | 143 | 0 | 56 | 23 | 29 | 0 | 10 | 0 | 0 | 0 | 9 | 0 | 0 |
| 47 | 38 | 71 | 51 | 31 | 21 | 0 | 13 | 0 | 0 | 28 | 0 | 65 | 0 |
| 183 | 115 | 0 | 364 | 90 | 67 | 73 | 74 | 59 | 71 | 105 | 0 | 45 | 0 |
| 38 | 0 | 0 | 50 | 49 | 0 | 0 | 52 | 56 | 0 | 0 | 17 | 37 | 37 |
| 42 | 41 | 45 | 57 | 56 | 50 | 0 | 24 | 0 | 0 | 0 | 0 | 0 | 0 |
| 83 | 0 | 0 | 125 | 83 | 0 | 0 | 83 | 65 | 0 | 0 | 0 | 0 | 0 |
| 75 | 64 | 0 | 34 | 77 | 5 | 0 | 172 | 267 | 34 | 118 | 26 | 0 | 0 |
| 340 | 121 | 91 | 40 | 59 | 194 | 150 | 81 | 26 | 57 | 69 | 45 | 52 | 34 |
| 129 | 139 | 182 | 100 | 37 | 26 | 0 | 30 | 65 | 0 | 0 | 0 | 0 | 0 |
| 75 | 50 | 59 | 23 | 19 | 80 | 95 | 82 | 100 | 59 | 0 | 0 | 0 | 0 |
| 286 | 158 | 38 | 24 | 100 | 12 | 12 | 148 | 167 | 0 | 0 | 0 | 0 | 0 |
| 194 | 115 | 214 | 0 | 154 | 40 | 0 | 67 | 0 | 31 | 24 | 32 | 23 | 0 |
| 105 | 174 | 77 | 42 | 67 | 0 | 0 | 238 | 188 | 59 | 125 | 125 | 48 | 83 |
| 77 | 263 | 364 | 273 | 235 | 61 | 136 | 56 | 48 | 78 | 68 | 33 | 89 | 74 |
| 229 | 71 | 79 | 43 | 31 | 48 | 11 | 64 | 222 | 53 | 111 | 83 | 40 | 0 |
| 77 | 118 | 50 | 100 | 200 | 77 | 77 | 26 | 34 | 250 | 273 | 50 | 15 | 31 |
| 70 | 44 | 0 | 55 | 43 | 100 | 0 | 121 | 0 | 0 | 0 | 0 | 0 | 0 |
| 154 | 28 | 19 | 211 | 38 | 217 | 91 | 143 | 0 | 0 | 0 | 0 | 0 | 0 |
| 185 | 206 | 167 | 263 | 65 | 74 | 48 | 68 | 143 | 79 | 83 | 54 | 0 | 0 |
| 30 | 63 | 0 | 0 | 26 | 0 | 0 | 0 | 56 | 0 | 29 | 50 | 19 | 23 |
| 191 | 207 | 267 | 75 | 48 | 128 | 59 | 86 | 57 | 158 | 45 | 87 | 69 | 33 |
| 300 | 286 | 0 | 25 | 26 | 0 | 0 | 11 | 36 | 27 | 22 | 8 | 0 | 0 |
| 267 | 242 | 115 | 333 | 120 | 259 | 73 | 63 | 167 | 167 | 100 | 31 | 18 | 0 |
| 95 | 81 | 182 | 91 | 69 | 91 | 0 | 104 | 0 | 0 | 34 | 86 | 14 | 115 |
| 70 | 182 | 0 | 87 | 167 | 41 | 67 | 106 | 71 | 200 | 47 | 0 | 30 | 91 |
| 241 | 154 | 100 | 91 | 200 | 0 | 42 | 225 | 278 | 250 | 15 | 44 | 60 | 111 |
| 116 | 89 | 0 | 27 | 23 | 10 | 26 | 207 | 167 | 154 | 19 | 56 | 125 | 0 |
| 220 | 80 | 45 | 141 | 103 | 0 | 129 | 171 | 100 | 93 | 82 | 156 | 102 | 43 |
| 50 | 120 | 100 | 40 | 68 | 59 | 0 | 98 | 100 | 154 | 133 | 136 | 74 | 0 |
| 50 | 129 | 0 | 200 | 0 | 16 | 0 | 167 | 68 | 125 | 48 | 31 | 53 | 0 |
| 119 | 280 | 0 | 62 | 53 | 0 | 74 | 19 | 0 | 0 | 29 | 0 | 0 | 0 |
| 200 | 48 | 158 | 75 | 0 | 133 | 60 | 192 | 56 | 333 | 83 | 67 | 0 | 28 |
| 107 | 82 | 53 | 35 | 30 | 0 | 0 | 97 | 97 | 0 | 19 | 42 | 0 | 0 |
| 130 | 33 | 0 | 47 | 44 | 61 | 31 | 50 | 56 | 40 | 36 | 0 | 0 | 0 |
| 83 | 157 | 300 | 46 | 100 | 67 | 0 | 61 | 94 | 91 | 61 | 65 | 188 | 357 |
| 87 | 65 | 87 | 37 | 81 | 81 | 25 | 33 | 35 | 21 | 13 | 22 | 19 | 7 |
| 24 | 87 | 48 | 70 | 87 | 26 | 71 | 48 | 45 | 42 | 0 | 0 | 24 | 0 |
| 213 | 200 | 63 | 211 | 52 | 32 | 154 | 16 | 0 | 0 | 36 | 0 | 0 | 0 |
| 176 | 308 | 143 | 79 | 269 | 83 | 40 | 35 | 133 | 0 | 0 | 0 | 0 | 0 |
| 320 | 250 | 200 | 0 | 43 | 24 | 0 | 133 | 63 | 0 | 83 | 111 | 100 | 67 |
| 98 | 51 | 125 | 138 | 11 | 4 | 108 | 163 | 150 | 125 | 27 | 125 | 39 | 75 |
| 9 | 30 | 33 | 27 | 31 | 0 | 0 | 43 | 200 | 92 | 14 | 57 | 24 | 78 |
| 68 | 24 | 86 | 75 | 33 | 48 | 0 | 156 | 231 | 167 | 136 | 98 | 91 | 77 |
| 205 | 190 | 222 | 54 | 37 | 17 | 67 | 28 | 143 | 0 | 0 | 53 | 0 | 0 |
| 324 | 60 | 91 | 74 | 74 | 61 | 100 | 200 | 158 | 91 | 136 | 71 | 43 | 0 |
| 82 | 67 | 320 | 222 | 39 | 34 | 31 | 86 | 57 | 158 | 45 | 87 | 69 | 33 |
| 64 | 102 | 167 | 118 | 95 | 143 | 34 | 51 | 200 | 308 | 182 | 59 | 0 | 0 |
| 107 | 300 | 0 | 222 | 65 | 48 | 77 | 64 | 71 | 56 | 0 | 23 | 18 | 25 |
| 57 | 68 | 167 | 231 | 261 | 48 | 313 | 105 | 94 | 48 | 111 | 118 | 121 | 71 |
| 163 | 150 | 75 | 13 | 16 | 18 | 0 | 47 | 100 | 0 | 0 | 24 | 0 | 0 |
| 95 | 24 | 37 | 0 | 83 | 0 | 0 | 61 | 83 | 0 | 20 | 0 | 11 | 0 |
| 261 | 167 | 111 | 216 | 125 | 43 | 53 | 105 | 77 | 0 | 45 | 0 | 0 | 25 |
| 200 | 125 | 73 | 0 | 184 | 41 | 125 | 45 | 103 | 214 | 87 | 80 | 80 | 63 |
| 80 | 105 | 0 | 111 | 80 | 100 | 107 | 63 | 91 | 273 | 74 | 26 | 13 | 53 |
| | | | | | | | 194 | 133 | 222 | 59 | 38 | 41 | 87 |
| | | | | | | | 64 | 240 | 69 | 36 | 0 | 10 | 24 |
| | | | | | | | 29 | 95 | 0 | 0 | 0 | 45 | 0 |
| | | | | | | | 97 | 21 | 91 | 0 | 24 | 11 | 0 |
| | | | | | | | 130 | 107 | 34 | 62 | 20 | 12 | 23 |
| | | | | | | | 115 | 67 | 29 | 28 | 81 | 0 | 0 |
| | | | | | | | 57 | 34 | 33 | 0 | 30 | 44 | 0 |
| | | | | | | | 33 | 121 | 63 | 111 | 0 | 36 | 0 |
| | | | | | | | 17 | 43 | 0 | 57 | 68 | 78 | 31 |
| | | | | | | | 74 | 250 | 91 | 0 | 95 | 0 | 0 |
| | | | | | | | 51 | 91 | 182 | 91 | 28 | 0 | 0 |
| | | | | | | | 61 | 37 | 133 | 71 | 26 | 0 | 0 |

| 52 | 53 | 43 | 0 | 74 | 0 | 0 |
| 87 | 36 | 0 | 9 | 43 | 9 | 13 |
| 188 | 68 | 15 | 58 | 104 | 48 | 63 |